\documentclass[aps,floatfix,twocolumn]{revtex4}

\newcommand{\be}{\begin{equation}}
\newcommand{\ee}{\end{equation}}
\newcommand{\bea}{\begin{eqnarray}}
\newcommand{\eea}{\end{eqnarray}}
 \renewcommand{\P}{{\cal P}}

 \newcommand{\fNL}{f_{\rm NL}}
 \newcommand{\unl}{\underline}

\def\c{\chi}
\def\s{\sigma}
\def\mc{m_\chi}

\def\e{\epsilon}
\def\d{\delta}
\def\p{\phi}

\begin{document}
\title{Tilted Ekpyrosis}

\author{Jos\'e Fonseca \& David Wands}

\affiliation{Institute of Cosmology \& Gravitation, University of Portsmouth, Dennis Sciama Building, Burnaby Road
Portsmouth, PO1~3FX, United Kingdom}
\begin{abstract}
We consider a simple model of cosmological collapse driven by canonical fields with exponential potentials. We generalise the two-field ekpyrotic collapse to consider non-orthogonal or tilted potentials and give the general condition for isocurvature field fluctuations to have a scale-invariant spectrum in this model. In particular we show that tilted potentials allow for a slightly red spectrum of perturbations as required by current observations. However a red spectrum of fluctuations implies that the two-field ekpyrotic phase must have a finite duration and requires a preceding phase which sets the initial conditions for what otherwise appears to be a fine-tuned trajectory in the phase space.
\end{abstract}
\maketitle

\paragraph*{\textbf{Introduction}}

Understanding the origin of structure in our Universe is one of the biggest challenges in modern cosmology. An inflationary expansion in the very early universe has become the standard explanation, addressing the flatness and the horizon problems as well as seeding an almost scale-invariant, nearly Gaussian distribution of inhomogeneous perturbations about a Friedmann-Robertson-Walker spacetime \cite{Linde:2005ht,Lyth:2009zz}. Nonetheless, it is interesting to ask if there are alternative scenarios that can source primordial perturbations consistent with current observations. We require primordial density perturbations which are well-described by a power spectrum $\P_\zeta(k)\propto k^{n_\zeta-1}$ where $0.944<n_\zeta<0.992$ \cite{Komatsu:2010fb}, and the distribution must be sufficiently Gaussian, such that the amplitude of the bispectrum with respect to the square of the power spectrum, given by the non-linearity parameter $\fNL$, is constrained to be $-10<\fNL<74$ \cite{Komatsu:2010fb} for local-type non-Gaussianity \cite{Wands:2010af}.

Pre-Big Bang models offer a possible alternative where the comoving Hubble-horizon shrinks during a collapse phase, generating a distribution of classical fluctuations on super-Hubble scales \cite{Gasperini:1992em,Wands:2008tv}. One of such model is an ekpyrotic collapse prior to the Big Bang \cite{Khoury:2001wf,Kallosh:2001ai,Lehners:2008vx} where a canonical scalar field with a steep, negative potential energy drives the contraction. The potential for this field is taken to be $V(\p)=-V_0\exp(-c\p)$ with $c^2\gg1$, which has a scale-invariant form, leading to a power-law collapse ($a\propto (-t)^p$ where $p=2/c^2$) and a power-law power spectrum of fluctuations. All collapse models face a challenge to connect this runaway collapse to a decelerated expansion, but in any case this single-field model predicts a steep blue spectrum of adiabatic density perturbations,  $n_{\zeta}\simeq3$ \cite{Lyth:2001pf}, in contradiction with observations.

An almost scale-invariant distribution of perturbations can be realised in the new ekpyrotic scenario \cite{Lehners:2007ac,Buchbinder:2007ad,Creminelli:2007aq} by considering a multi-field system \cite{Finelli:2002we}. Each field has its own steep, negative potential, e.g., $V(\phi_1,\phi_2)=-V_1e^{-c_1\phi_1}-V_2e^{-c_2\phi_2}$. One can perform a rotation in field space and define an adiabatic direction, $\sigma$, and an isocurvature direction ($\chi$)~\cite{Malik:1998gy,Gordon:2000hv,Finelli:2002we,Koyama:2007mg}, i.e.,
\be
 \label{newV}
V(\sigma,\chi)=-e^{-c\sigma}\left(V_1e^{-(c_1/c_2)c\chi}+V_2e^{(c_2/c_1)c\chi}\right) \,,
\ee
with $c^{-2}=\sum_ic_i^{-2}$. A power-law solution exists where the field $\chi$ sits in the extremum of the potential, $\chi_0$, yielding a power-law solution driven by the exponential dependence, $V\propto e^{-c\sigma}$. The adiabatic fluctuations have a steep blue spectrum as before, but isocurvature fluctuations can also source the primordial density perturbation. The isocurvature field spectral tilt is given by $n_\chi-1\simeq 4/{c^2}$ during an ekpyrotic contraction with $c^2\gg1$; therefore the power spectrum can be nearly scale invariant.

An essential feature of this two-field model is that the the power-law solution solution with $p=2/c^2$ is unstable; there is a tachyonic instability since the effective mass-squared of the $\chi$ field is negative \cite{Koyama:2007mg}.
Such an instability is necessary to achieve an almost scale-invariant spectrum.
Quantum fluctuations on the Hubble scale have a power spectrum $\mathcal{P}_{\chi}\simeq({c^4}/4)\left({H}/{2\pi}\right)^2$ which grows rapidly during collapse, therefore the power spectrum on larger scales must also experience a rapid growth, proportional to $H^2$, in order to keep pace with the growing power on the shrinking Hubble scale.

This raises the question of how the universe started sufficiently close to this unstable solution, which we will return to later. However the tachyonic instability does provide a mechanism to convert isocurvature field fluctuations into density perturbations \cite{Koyama:2007ag}. The growth of the $\chi$ field leads to a change from the two-field solution with $p=2/c^2$ to a single-field solution with either $p=2/c_1^2$ or $p=2/c_2^2$. The corresponding change in the local equation of state, controlled by the local value of the $\chi$ field, leads to a density perturbation, $\zeta\propto\delta\chi$ \cite{Koyama:2007ag}.
Other mechanisms have also been proposed which could convert the isocurvature field fluctuations to density perturbations including a kinetic conversion due to an abrupt change in the field trajectory after the ekpyrotic phase \cite{Lehners:2007ac} or a curvaton-type conversion due to modulated reheating in an expanding phase following the bounce \cite{Battefeld:2007st}. In any case any linear process preserves the scale dependence of the power spectrum and we have $n_\zeta=n_\chi$. Note however that the power spectrum is slightly blue, $n_\zeta>1$, in tension with current observations \cite{Komatsu:2010fb}.

Non-linearity in the evolution of perturbations also provides important constraints on the model. The tachyonic conversion of isocurvature field fluctuations into density perturbations leads to local-type non-Gaussianity characterised by the nonlinearity parameter \cite{Koyama:2007if} $\fNL=-(5/12) c_I^2$ for $I=1,2$. Given that we must have $c_I^2>c^2$ this implies $\fNL<(-5/3)(1-n_{\zeta})^{-1}$, e.g., if $1-n_{\zeta}<0.01$ we require $\fNL<-100$, in contradiction with observations. Alternative conversion processes can lead to model-dependent results for non-Gaussianity and in particular the kinetic conversion can lead to $\fNL\sim\pm c$ which may be compatible with observational constraints given above.

In this letter we will look at consequences of simple generalizations of the new ekpyrotic scenario to include non-orthogonal potentials and how this alters the predicted distribution of super-Hubble perturbations and the problem of initial conditions.

\paragraph*{\textbf{Tilting the potentials}}

We will consider $n$ canonical scalar fields, $\unl{\p}=(\p_1,\ldots,\p_n)$, with $m$ exponential potentials
\be
 \label{Vphi}
V(\unl{\phi})=-\sum_{J=1}^mV_Je^{-\unl{c}_J.\unl{\phi}}
\ee
where $\unl c_J=(c_{J1},\ldots,c_{Jn})$.
We recover the new ekpyrotic model, described above, as a special case of two orthogonal vectors $\unl c_1.\unl c_2 = 0$, but in the following we will consider the more general case of non-orthogonal or ``tilted'' potentials, such that $\unl c_I.\unl c_J\neq 0$ \cite{Finelli:2002we}.
We restrict our discussion to $V_J>0$ so that every term in Eq.~(\ref{Vphi}) is negative and $V<0$. The case of positive potentials, $V>0$, was discussed previously in the context of assisted inflation \cite{Copeland:1999cs,Hartong:2006rt}. We will assume that the $m$ different vectors, $\unl c_J$, are linearly independent. Hence our analysis is also restricted to $m\leq n$ and we assume that the fields are not trapped, so that there always exists a regime with finite energy density in which $V_Je^{-\unl{c}_J.\unl{\p}}\to0$ for any given potential.

We note that we could choose to work with fields $\varphi_J\propto \unl{c}_J.\unl{\phi}$ aligned with the potentials in Eq.~(\ref{Vphi}) but then these fields would have a non-diagonal metric in field space, i.e., be non-orthogonal for $\unl c_I.\unl c_J\neq 0$.

The evolution equation for the canonical fields is given by
\be \label{evolueq}
\ddot{\unl\phi}+3H\dot{\unl\phi}+\sum_J\unl c_JV_Je^{-\unl{c}_J.\underline{\phi}}=0
\ee
where the Friedmann equation for $H\equiv\dot a/a$ is
\be \label{friedeq}
H^2=\frac13\left(\frac12\dot{\unl{\p}}.\unl{\dot\p}-\sum_JV_Je^{-\unl{c_J}.\unl{\p}}\right)
\ee
We have set $8\pi G=1$ and dots correspond to derivatives with respect to cosmic time.

Firstly let's look at the stability of this type of system. To do so let's follow \cite{Copeland:1997et,Koyama:2007mg} and define
\bea
x_i&=&\frac{\dot\p_i}{\sqrt6H}\,,\\
y_J&=&\frac{\sqrt{V_Je^{-\unl{c_J}.\unl{\p}}}}{\sqrt3H}\,.
\eea
Using Eqs (\ref{evolueq}) and (\ref{friedeq}) one finds
\bea
\frac{dx_i}{dN}&=&-3x_i\left(1-\sum_kx_k^2\right)-\sqrt{\frac32}\sum_Jc_{Ji}y_J^2\\
\frac{dy_J}{dN}&=&y_J\sum_ix_i\left(3x_i-\sqrt{\frac32}c_{Ji}\right)
\eea
We can then study fixed points corresponding to scaling solutions \cite{Copeland:1997et,Tolley:2007nq,Hartong:2006rt,Heard:2002dr}. They are given by $dx_i/dN=dy_J/dN=0$. We find the following fixed points:

\begin{enumerate}

\item{Zero-potential fixed point}

These points are characterized by $y_J=0, \forall ~J $ and $\unl x.\unl x=X^2=\sum_ix_i^2=1$. This kinetic energy-dominated collapse with $a\propto (-t)^{1/3}$ is unstable whenever there exists at least one potential with $c_J^2>6$.

\item{Single-potential fixed point}

In the case where only one potential is non-zero, i.e., $y_K\neq0$ while $y_J=0, \forall ~ J\neq K$, we then have a fixed point
\be
y_K=\sqrt{\frac{c_K^2}6-1} \qquad x_i=\frac{c_{Ki}}{\sqrt6} \,,
\ee
where $c_K^2=\sum_ic_{Ki}^2$ and we require $c_K^2>6$. This corresponds to a power law solution of the scale factor with $a\propto(-t)^p$, where $p=2/c^2_K$. This collapse is stable with respect to the zero-potential solution.

\item{Double-potential fixed point}

In the case where two potentials are non-zero, $y_K\neq0$ and $y_L\neq 0$, we have a fixed point where
\be
x_i=\frac{c_i}{\sqrt6} \,,
\ee
and
\be
y_K=\sqrt{\frac{c_L^2-\unl{c_L}.\unl{c_K}}{\left|\unl{c_L}-\unl{c_K}\right|^2}\left(\frac{c^2}6-1\right)}\,,
\ee
with
\be
 \label{2c}
\unl{c}=\frac{\left(c_K^2-\unl{c_L}.\unl{c_K}\right)\unl{c_L}+\left(c_L^2-\unl{c_L}.\unl{c_K}\right)\unl{c_K}}{\left|\unl{c_L}-\unl{c_K}\right|^2}\,.
\ee
This corresponds to a power-law solution of the scale factor with $p=2/c^2$ where $c^2=c_K^2c_L^2-\left(\unl c_K.\unl c_L\right)^2$.
For this solution to exist we require $c^2>6$ and $\unl c_K.\unl c_L<{\rm min}\{c_K^2,c_L^2\}$, and in this case this double-potential collapse is always stable with respect to kinetic-dominated collapse but unstable with respect to single-potential collapse.

\item{Multiple-potential fixed points}

The scaling solution for two potentials can be generalised to the case of multiple tilted potentials. We again find $x_i=c_i/\sqrt{6}$ where we have
\be
 \label{multicvector}
 \unl{c} = \frac{\sum_I \left(M^{-1}\right)_{IJ} \unl{c}_J}{\sum_{I,J} \left(M^{-1}\right)_{IJ}} \,,
 \ee
where we define the matrix $M_{IJ}\equiv\unl{c}_I.\unl{c}_J$ and $(M^{-1})_{IJ}$ is its inverse. Hence we have a power-law solution with $p=2/c^2$ where \cite{Copeland:1999cs}
\be
 \label{multic}
 c^2 = \left( \sum_{I,J} \left(M^{-1}\right)_{IJ} \right)^{-1} \,.
 \ee

\end{enumerate}

A system with many exponential potentials can have many different fixed points. $n$ fields with $m\leq n$ potentials of the form given in Eq.~(\ref{Vphi}) with independent $\unl{c}_J$ will have $2^m-1$ different fixed points with at least one non-zero potential. For instance, if we have 3 potentials there will be one scaling solution with 3 non-zero potentials ($y_K\neq0$), three scaling solutions with 2 non-zero potentials, and three fixed points each with a single non-zero potential. In each case we can use the general result for the multiple-potential fixed point given in Eqs.~(\ref{multicvector}) and (\ref{multic}) where the sums are to be taken over the non-zero potentials. This reduces to Eq.(\ref{2c}) for two tilted potentials, or $c^{-2}=\sum_I c_I^{-2}$ for multiple orthogonal potentials.

\paragraph*{\textbf{Tilted ekpyrosis}}

Let's consider a new ekpyrotic scenario with non-orthogonal potentials. From now on we will discuss the case when we have two potentials and two fields. Without further loss of generality we set
\bea
\unl{c_1}&=&c_1(1,0)\\
\unl{c_2}&=&c_2(\sin\theta,\cos\theta)\,.
\eea
We recover the case of orthogonal potentials in the limit $\sin\theta\to0$.
The rotation to adiabatic and isocurvature fields in field space is given by
\bea
\sigma&=&\frac{(c_2\cos\theta)\phi_1+\left(c_1-c_2\sin\theta\right)\phi_2}\Delta, \\
\c&=&\frac{\left(c_1-c_2\sin\theta\right)\phi_1-(c_2\cos\theta)\phi_2}\Delta
\eea
where $\Delta^2\equiv c_1^2-2c_1c_2\sin\theta + c_2^2$. The potential in terms of the adiabatic field $\s$ and the isocurvature field $\c$ is $V(\s,\c)=-e^{-c\s}U(\c)$ where
 \be
 c^2 = \frac{c_1^2c_2^2\cos^2\theta}{c_1^2+c_2^2-2c_1c_2\sin\theta} \,,
 \ee
and
\bea
U(\c)=&V_1\exp\left[-\frac{(c_1-c_2\sin\theta)c}{c_2\cos\theta}\c\right]+&\nonumber\\
&+V_2\exp\left[\frac{(c_2-c_1\sin\theta)c}{c_1\cos\theta}\c\right]&\,.
\eea
which reduces to (\ref{newV}) when $\theta=0$.
We note that $V_1,V_2>0$ so $U(\c)$ is bounded from below and has a minimum at $\chi=\chi_0$ for $c_1c_2\sin\theta<{\rm min}\{c_1^2,c_2^2\}$. Thus there is an ekpyrotic, power-law solution with $\c=\c_0$ and $V\propto e^{-c\sigma}$.
Around the minimum we can expand $U(\c)$ as
\be
 \label{Uchi}
U(\c)\simeq U_0\left(1+\frac{\mu_\c^2}2(\c-\c_0)^2+\ldots\right)
\ee
where
\bea
\mu_\c^2&=&c^2-c_1c_2\sin\theta > 0\,.
\eea

Quantum fluctuations in the adiabatic field, $\sigma$, lead to a steep blue spectrum of density perturbations, $n_\zeta=3$ in the fast-roll limit\cite{Lyth:2001pf}.
A nearly scale-invariant spectrum of perturbations can instead originate from isocurvature fluctuations in the isocurvature field, $\c$.

Linear perturbations in the isocurvature field obey
\be \label{evolvedeltachit}
\ddot{\delta\chi}+3H\dot{\delta\chi}+\left(\frac{k^2}{a^2}-\frac{2\mu_\c^2(c^2-6)}{c^4t^2}\right)\delta\chi=0\,.
\ee
It is convenient to work in conformal time defined by $dt=ad\tau$. Using the Bunch-Davies vacuum state to normalize the amplitude of quantum fluctuations at early times ($k\tau\rightarrow -\infty$) one finds
 \be
 a\delta\chi(\tau)=\frac{\sqrt{\pi}}{2} {e^{-i\frac\pi2(\nu+\frac12)}}(-\tau)^{1/2}H_\nu^{(2)}(-k\tau)
 \ee
where $H_\nu^{(2)}(-k\tau)$ is a Hankel function of the second kind and
 \be
 \nu^2=\frac94 -\frac{{2c^4-2(3+\mu_\c^2)c^2+12\mu_\c^2}}{(c^2-2)^2} \,.
 \ee

The scale dependence of the power spectrum is given by $n_{\chi}-1\equiv {d\ln \P_{\chi}}/{d\ln k}=3-2\nu$.
To obtain an exactly scale-invariant spectrum we require $\nu=3/2$ and hence
 \be
 c_1 c_2 \sin\theta = -\frac{3c^2}{c^2-6} \,.
 \ee
Note that, as expected, a scale-invariant spectrum requires that the two-potential collapse is an unstable point in the phase space.

In the fast-roll limit $c^2\gg1$ we have
\be
\nu^2\simeq\frac 14 +\frac{2(3+\mu_\c^2)c^2-12\mu_\c^2}{c^4} \,.
\ee
%
We will focus on the fast roll, $c^2\gg1$, and small angle case, $\theta\ll1$.
In this case
\be
n_{\chi} - 1 \simeq
 \frac4{c^2}\left(1+\frac{c_1c_2}3\theta\right) \,.
\ee
So we find a nearly scale-invariant red spectrum if $\theta\lesssim-3/c_1c_2$.
The relative tilt of the potentials in field space, $\theta$, alters the effective mass of the isocurvature field enabling us to obtain a slightly red spectrum in contrast to the case of orthogonal potentials.

Another important constraint comes from the observed Gaussian distribution of primordial perturbations. Due to the steep exponential potentials, the isocurvature field fluctuations do not remain Gaussian on super-Hubble scales. The resulting non-Gaussianity of the subsequent density perturbations depends on the conversion process. In the simplest case the $\chi$ field perturbations away from the two-potential fixed point lead to a tachyonic transition to the single-potential solution.

Assuming there is an instantaneous transition from two-potential scaling solution to single-potential solution, when the Hubble rate reaches $H=H_T$ then the integrated expansion from $H_i$ before the transition to $H_f$ after the transition is given by $N=-2\left[\frac{c_1\sin\theta-c_2}{c_1c_2\cos\theta}\right]^2 \ln |H_T| +\rm constant$.
Following Koyama et al \cite{Koyama:2007if} we can compute the non-linear growth of the isocurvature field and the dependence of $H_T$ upon the initial field fluctuations to determine the non-linear expansion, $N$, and hence the bispectrum.
In the fast roll limit we again recover $\fNL=-(5/12)c_1^2<-(5/12)c^2$. Although $c$ is no longer uniquely determined by the tilt of the power spectrum, it is nonetheless required to be large in the fast-roll limit. On the other hand, if we require $\fNL>-10$ this demands $6<c^2<24$.

\paragraph*{\textbf{Initial conditions for a red tilt}}

Having successfully obtained a red spectrum of density perturbations by tilting the potentials we have also created a new problem of initial conditions for the scenario as we will now demonstrate.

Although a classical solution can spend an arbitrarily long period of time close to the unstable double-potential fixed point, quantum fluctuations in the isocurvature field inevitably lead to a tachyonic transition within a finite time. The transition occurs when $\mu_\c^2\langle\delta\c^2\rangle\sim 1$, where
\be
 \langle \delta\c^2 \rangle = \int {\cal P}_\c(k)d\ln k \,,
  \ee
where 
the integral is over all wavenumbers for which we can treat the fluctuations as effectively classical, usually taken to be all super-Hubble scales. For a blue spectrum of perturbations the integral is dominated by the shortest wavelengths, i.e., the Hubble scale, and we have $\langle\delta\c^2\rangle \simeq (c^4/4)\left({H}/{2\pi}\right)^2$. Thus, taking $\mu_\c^2\simeq c^2$, the transition must occur when $|H|\lesssim c^{-3}$, but it can begin arbitrarily far in the past.

But for a red spectrum of fluctuations in the isocurvature field, the variance of the field is dominated by the longest wavelengths. If the new ekpyrotic phase started infintely far in the past, then the variance of the field on the largest scales would be infinitely large leading to a contradiction as we require the field to be close to the double-potential fixed point. Thus the new ekpyrotic phase must have started a finite time in the past and we have
\be
 \mu_\c^2 \langle \delta\c^2 \rangle \simeq \frac{c^6}{4} \left(\frac{H}{2\pi}\right)^2 \left( \frac{k_*}{k_i} \right)^{1-n} \,.
 \ee
where $k_*$ is the comoving Hubble scale and $k_i$ is the initial comoving Hubble scale at the beginning of the new ekpyrotic phase.
Assuming we have a slightly red spectrum with $1-n\simeq 0.01$ and requiring a new ekpyrotic phase which lasts at least 10 e-folds, i.e., $k_*/k_i>e^{10}$, we require that the transition completes when $|H|<(1-n)^{1/2}c^{-3}$.

We require a phase preceding the new ekpyrotic phase which sets the classical background field sufficiently close to the fixed point, and ensures that the isocurvature field has a sufficiently small variance on large scales at the start of the new ekpyrotic phase. There are several possibilities, one being that the isocurvature field has a mass parameter that changes during the evolution, inserting $\mu_\chi^2(\sigma)$ in Eq.(\ref{Uchi}). This could both stabilise $\chi=\chi_0$ at early times and offers another way produce a red spectrum at late times \cite{Buchbinder:2007tw}. However such a potential cannot be realised within the context of simple exponential potentials (\ref{Vphi}), and lies outside the class of scale-invariant potentials with scaling solutions \cite{Tolley:2007nq}. We expect that a time-dependent $\mu_\chi^2(\sigma)$ would lead to a running of the tilt, $n_\zeta(k)$ and a scale-dependent  non-Gaussianity, $\fNL(k)$.

Within our simple model (\ref{Vphi}) one possibility could be a preceding phase described by a multiple-potential scaling solution that itself would be unstable with respect to the two-potential solution. But the spectrum of the isocurvature perturbations requires a careful analysis of the three (or more) potential system, and could be highly model-dependent. An alternative preceding fixed point already present in our two field model is the kinetic fixed point with vanishing potential energy. This is an unstable fixed point but it does describe the generic behaviour of the system as $t\to-\infty$.

The kinetic-dominated fixed point where the potentials are negligible is in fact the basis of the pre-big bang models proposed by Gasperini and Veneziano \cite{Gasperini:1992em}. It is well-known that a kinetic-dominated collapse leads to a steep blue spectrum of perturbations for any massless fields. Thus the isocurvature field naturally has negligible perturbations on large scales. On the other hand a priori there seems no particular reason why the classical background trajectories should approach close to the new ekpyrotic (double-potential) solution which is a saddle point in the phase-space \cite{Koyama:2007mg} rather than proceding directly to the old ekpyrotic (single-potential) solutions which are the stable late-time attractors.

Finally we note that in principle we might disregard the ensemble average for $\langle\chi^2\rangle$ on large scales and assume that simply by chance quantum fluctuations away from $\chi=\chi_0$ in our local patch are unusually small. This is unlikely a priori but one might appeal to some anthropic argument that only these regions are capable of giving rise to observers \cite{Lehners:2011ig}.

\paragraph*{\textbf{Conclusions}}

In this letter we have shown that an ekpyrotic collapse driven by two scalar fields with non-orthogonal potentials can give a scale-invariant or slightly red tilted spectrum of perturbations. This is in contrast to the original ekpyrotic collapse with a single field \cite{Khoury:2001wf} which produces a steep blue spectrum \cite{Lyth:2001pf}, or {\em new} ekpyrotic collapse with two orthogonal potentials \cite{Lehners:2007ac,Buchbinder:2007ad,Creminelli:2007aq} which yields an almost scale-invariant, but slightly blue spectrum of perturbations. To obtain a slightly red spectrum we fine-tune the tilt such that the angle $\theta\sim 0.01$ in the fast-roll limit.

This two-potential solution is an unstable saddle point in the phase-space and a red tilted spectrum of tachyonic field fluctuations can therefore only exist over a finite range of scales. Thus the two-potential solution can only exist for a finite time. This is possible for a particular class of solutions in the phase space which must evolve from a kinetic-dominated initial state to approach sufficently close to the two-potential saddle point. The late-time attractor in the phase space is a single-potential-dominated collapse, i.e., the old ekpyrotic collapse \cite{Khoury:2001wf}.

If the tachyonic transition from two-potential to single-potential collapse occurs then this naturally converts the isocurvature field fluctuations into density perturbations. However this potentially leads to a large, negative non-Gaussianity parameter, $\fNL<-(5/12)c^2$, in the fast-roll limit, $c^2\gg1$, in contradiction to the observations.

By studying a simple two-field system we have a well-defined model within which we can calculate the quantum field perturbations about classical trajectories during a cosmological collapse. However it leaves unanswered the question of whether the required tilted potentials can be realised within a string theory setting, as originally envisaged in the ekpyrotic scenario \cite{Khoury:2001wf}, or how the initial state evolves sufficiently close to an unstable saddle point in the phase space. This would require a preceding phase \cite{Buchbinder:2007tw}, such as is envisaged within the cyclic scenario \cite{Steinhardt:2001st}.
In all these scenarios we still need to understand whether, and if so how, the universe emerges from collapse to begin expanding and recover the standard hot big bang.

\acknowledgements

The authors are grateful to Kazuya Koyama, Shuntaro Mizuno and Filippo Vernizzi for helpful discussions. JF is supported by FCT (Portugal) fellowship SFRH/BD/40150/2007. DW is supported by STFC grant ST/H002774/1.

\end{document}